# A Synthesis and a Practical Approach to Complex Systems

Nicolas Brodu

Department of Computer Science and Software Engineering
Concordia University, Montreal, Quebec, Canada, H3G 1M8

Current Affiliation : INRIA Rennes Bretagne Atlantique, 35042, France

Tel: +33 2 99 84 71 51, E-mail: nicolas.brodu@free.fr

**Abstract-** This document is both a synthesis of current notions about complex systems, and a practical approach description. A disambiguation is proposed and exposes possible reasons for controversies related to causation and emergence. Theoretical considerations about simulations are presented. A justification is then given for the development of practical tools and techniques for the investigation of complex systems. A methodology for the usage of these tools is finally suggested, illustrated by application examples.

## 1 Introduction

What is a complex system, its main features and properties? What does it mean that something is emergent?

Since the advent of modern calculus in the 17$^{th}$ century with Newton and Leibnitz, the dominant philosophy has been that of integration: from the reasoning on an elementary scale we could sum up and obtain global results about a system (for example, movements of planets). Of course some equations describing a system behaviour cannot be integrated, so one cannot find a shortcut that allows a direct computation for a prediction at the higher scale. Even when such a shortcut exists it is not always applicable: Exponential relations for example were sensitive to initial conditions even before the discovery (reviewed by James Gleick [GLEI87]) of Chaos theory, which brought the notion of being locally exponentially divergent and globally bounded at the same time[1]. Yet, and especially with computers, approximate methods and numerical integrations were developed that can produce reasonable results, and they still form the majority of industrial simulations to date.

Another approach is the study of the high-level properties of the system, considering entities defined at a global scale scale, as discussed by Russ Abbott in [ABBO06]. Then one could try modelling these entities and their interactions directly rather than by applying the more traditional integration approach. Some other systems are self-similar at different scales and may be better analysed by yet another method as explained by Benoit Mandelbrot [MAND82]. There are also universal phenomena and global properties that may be observed whatever the underlying equations. So if we now look at the problem top-down any phenomenon that we observe at the system macro-scale but that we cannot somehow relate to micro-states poses a similar problem as before but the other way around.

The reasons why a phenomenon defined at a high level cannot be related to low-level properties may be multiple, from simple ignorance of hidden relations to theoretical uncomputability. But whatever these reasons the same practical issue remains between the high-level scale and the underlying micro-scale elements: the phenomenon is then often labelled "emergent". The notion of emergence has progressed over time, and its history is reviewed by Peter A. Corning in [CORN02]. Refinements about possible reasons for the failure to relate micro and macro properties were proposed, but overall the same idea remains in one form or another.

These main ideas may depict a legitimate field of study, but the wide range of application domains they're supposedly applicable to makes it difficult to synthesize results into a consistent framework. Attempts at creating a theory of emergent phenomena often end up having to define concepts that are specific to that attempt. Consequently, there are as many definitions as frameworks, and no real common theory. And this document is thus not a proposal to create yet another framework.

Yet if any progress is to be made on complex systems a formalization is necessary at some point. See for example the proposals by Cosma R. Shalizi [SHAL01] and by Aleš Kubík [KUBI03]. As mentioned above, these mathematical frameworks do not encompass (to date) all aspects that were proposed by other definitions. What is called emergent by some is outside the definition of others. Too broad definitions are rejected because they are either inapplicable or they would include a range of phenomena that we intuitively do not label as "emergent"; while too restricted definitions miss one or another of such phenomena. It is likely that no definition of emergence may satisfyingly correspond to our intuition (this point is explained in Section 2.4.1), and conversely that any successful hypothetical theory on emergence would include counter-intuitive effects.

However there is no reason to think that Complexity Science cannot be handled by the traditional approach exemplified by Thomas S. Kuhn [KUHN62]: by using incremental steps, with predictive testing, refinement of the main concepts, that allow to validate or not the main

---

[1] See also http://cscs.umich.edu/~crshalizi/notebooks/chaos.html (checked on 2008/16/01)

theories, etc. This is what Kuhn calls *normal science*, posed as a necessary condition for further ground-breaking discoveries. Jochen Fromm also suggests in [FROM06] that the difficulty to study emergence in a system is equivalent to the difficulty to create a higher-level theory for that system, which is a direct application of *normal science* in the context of complex systems.

At this point what is thus essentially missing is the availability of adequate investigation tools: if a generic theory of complex systems is still out of reach, at least can we build the means to facilitate their analysis and the better characterization of the different processes involved in what we call emergent phenomena. In turn, we might then hope to achieve the desired global formalization step thanks to the result of our better understanding of complex systems. Fortunately we now also have computers and their associated formidable processing power. Some of the best investigation tools will thus take the form of practical algorithms and computer simulations.

The next section proposes a review of the domain and what are possibly the main sources of the controversies associated to emergence and complex systems. The goal is not to engage in a philosophical debate about the merits of such or such framework and take position. The goal is to identify what are the main notions and what they entail. This helps avoid controversies, and some well-defined basis is necessary for a sound future work. A contribution concerning computer simulations is also proposed at the end of section 2, concerning how such simulations are adequate for the study of complex systems.

Section 3 then presents an illustration of different possible approaches. The choice was made to concentrate on practical problems: prediction and control in complex systems. In particular, attention is given to top-down global control and bottom-up micro control. It is also explained why formalizing higher-levels as their own independent frameworks can help in increasing this control, as well as why it allows to make reasonable predictions about a system.

Finally, it is discussed how to extend these ideas, and a synthesis and generalization is done in section 4 for further work.

# 2 Review of emergence related concepts

This section reviews different concepts related to emergence. The main goal is to clarify these notions and to pose a framework for the further sections of this document, but this section is also the occasion to synthesize previous work and present an original comprehensive digest. The intent is to remain factual and not to engage in the many controversies surrounding these emergence related concepts. A more engaged theoretical discussion and contribution will be presented at the end of this section, concerning computer simulations. Justification will then be given for the choice of a practical approach for the investigation of emergent phenomena.

## 2.1 Review of common notions

Emergence does not mean anything in itself, so long as the concept is not clarified. "Deaf dialogues" may be engaged over whether a phenomenon is emergent or not, if both sides do not consider the same definition. This part is divided into three subsections: The basic constituents are presented in Section 2.1.1, they form the building blocks for the definitions of Section 2.1.2. These definitions are descriptive only: they may be used to clarify the domain by classifying and qualifying the properties of complexity and emergence, but they have no predictive value (so far). Quantifiable aspects of complexity are described in Section 2.1.3. These quantities are necessarily dependent on some formalization, hence become "reductionist" compared to the holistic concepts presented in the Sections 2.1.1 and 2.1.2. However, they do have a predictive value, and may thus form the most promising approach for a formalized theory of emergence.

### 2.1.1 Ingredients for a complex recipe

The notions presented in this subsection form a common basis for complex systems frameworks. These ideas are generic and applicable to many systems.

The idea there are **Levels of investigation** correspond to the intuitive notion that was presented in introduction: That a system can be studied at different scales, or at least at a micro-level and at a global level. When the process can be repeated for yet another level this defines a hierarchy. This notion is not new: Philip E. Agre [AGRE03] reviews and explains the static vs. dynamic hierarchies issue that was presented by Herb A. Simon in [SIMO69], related to how the levels are defined. Russ Abbott [ABBO06] also considers hierarchies and static and dynamic emergence, and these notions are presented in Section 2.1.2. Peter A. Corning [CORN02] proposes an historical perspective where older articles also convey the ideas of hierarchies, especially in the life sciences.

In either case there are observed entities made of smaller constituents, and some features of the entities are not easily linked (or reduced) to the constituents. See also the "whole and the sum of parts" next entry. The hierarchical organization of levels occurs when such entities form themselves the basic constituents of yet another larger entity, and so on. For example, a cell, an organ, an organism, a social organization, etc. In the particular context of life sciences John Maynard Smith and Eörs Szathmáry [SS95] explore explicitly the transitions from one level to the other.

Yet the boundary between levels is not always very clear, and some constituents may interact at different scales. Alternatively, scope, together with resolution and the states of a system were proposed by Alex J. Ryan in [RYAN06] as better notions: study should then be done on entities defined at their proper resolution in space and time, and whether there are other such entities at the same scale to form a "level" is irrelevant. In this view, the levels themselves could possibly be emergent properties.

While attractive, the scope/resolution approach does not solve the main issue of the relations between the components, irrespectively of how they are defined. Yet, Russ Abbott [ABBO06] in particular shows that reasoning on entities directly rather than on global levels solves a number of definition issues and thus clarifies the situation.

The notion that there are "levels" of investigation is a handy conceptual tool, but it is unfortunately defined precisely because it allows to pose the question of emergence (higher-level entities are said to emerge from a lower-level), thus forming a circular definition. Jaegwon Kim [KIM99] states that *a layered model [..] provides an essential framework needed to formulate the emergentist/reductionist debate.* The problem is also that the "layer" defined implicitly by one "emergent" entity may not correspond to the one of another. Hence there is no global layer but rather a continuum of scales with their own properties and entities, defined with respect to other entities at a lower or equal scale.

**The whole and the sum of parts** refer to the statement by Aristotle [CORN02] that both are not identical. By extension, this is the same idea as the one of synergy between components: a higher-level entity comprising lower-level elements is a "whole" that is not just the mere juxtaposition of these elements. A reductionist approach is that the "sum" in "sum of parts" is more complicated than a simple linear combination and thus explains our apparent inability to relate the whole with the parts. Then, since linear relations form only a small fraction of all possible relations[2], this explains the apparent universality of the "whole vs part" issue, though there is nothing special going on and the notion is better investigated on a case by case basis. On the holistic side [KAUF93], the parts are said to self-organise due to their relations, but there is also irreducibility of some higher-level function of the whole (consequently this functionalist view is not just a matter of non-linear relations). In any case, "this parts and whole" approach to emergence is perhaps historically [CORN02] the first approach, and it is still a topic of controversy. There are also complications with the notion of causality, which are detailed in Section 2.2.

**Interactions** between the elements must be taken into account, and they must be sufficiently complex so there can be a "whole" which is not just elements side by side. Interaction graphs and networks then define as much of the global "whole" as the elements own nature. Such networks then offer a connection with dynamical systems and graph theory. They can be simulated and their properties can be studied on a large scale (see Andrew Wuensche [WUEN02] and Réka Albert and Albert-László

---
2 Stanislaw Ulam compared non-linear mathematics to non-elephant zoology…

Barabási [AB02]). John Holland in [HOLL98] models the relations between the elements as **constrained generating procedures**. Similarly, when the parts can learn and adapt to their environment the system is called a **Complex Adaptive System** [HOLL98].

**Open dissipative** structures were initially defined in a thermodynamical framework by Grégoire Nicolis and Ilya Prigogine [NP77]. The idea may be extended: so long as the underlying assumptions allow for a definition of a generic notion of energy, and the system under consideration allows that energy to flow, then entities in that system may "use" this energy [ABBO06]. Extensions to this framework are when the entities can store energy and then use that reserve later in time [KAUF00], and when the entities simply use the energy to perpetuate themselves, which then leads to the notion of **autonomous structures** [ABBO06]. Other extensions consider how an autonomous structure may use the output flow in relation to its environment as a mean of action (with the corresponding form of causal relationship, see Section 2.2). The notion of **empowerment** by Klyubin *et al.* [KPN05] represents precisely this ability to act on the environment, but also relates it to the feedback the autonomous agent may get from its actions.

The notion of **energy** may be abstracted in a functionalist point of view. For example, in social contexts, energy may be related to available skills, money, or time; In artificial life contexts energy may be CPU execution slots; In discrete dynamical systems energy may be related to a system state change (and its dissipation would be the fusion of trajectories). Generally speaking, energy is functionally defined by the capacity of the entities in the system to use it. Of course this leads to a circular argument. Howard Pattee's semantic closure concept [PATT95] can also be used as a justification for a separation of the emergent level, when the usage of the energy has an intrinsic signification for the entities in the system (see also the semantic vs syntactic entry in Section 2.1.2). The notion of energy may then be used formally in the higher level.

**Self-organisation** is concerned with the internal structure of a system, and how that structure evolves without external intervention. [KAUF93] proposes that self-organisation is the result of positive feedback loops (see above). The term self-organisation is credited to William Ross Ashby [ASHB56] in a pioneer work on cybernetics, but the notion has now extended to a point where it is ubiquitous. Cosma Rohilla Shalizi presents an extensive effort [SHAL01] to clarify the notion in the context of time series, and equates self-organisation to a rise in statistical complexity (with a working data-based algorithm for computing this value, see the corresponding entry in 2.1.3). Another definition for self-organisation is the *state-space description* proposed by Francis Heylighen [HEYL01]: *self-organisation as the appearance of coherence or correlation between the system's components is equivalent to the reduction of entropy,* which is in some cases contradictory with the statistical complexity interpretation. There are other definitions, like the positive feedback loop approach previously mentioned. A generally applicable and consensual notion of self-organisation has thus yet to be defined.

**Hypercycles** [ES70] are another name for **positive feedback loops**, applied in a pre-biotic biological context. Proto-cells in the form of compartments allow different chemicals to concentrate and then react. The feedback loop is when the resultant of one reaction enhances the next, in this case RNA strings are translated to enzymes which catalyse the next reaction. This mechanism is a *natural principle of self-organisation* and an important hypothesis for the appearance of life [SS95]. Stuart A. Kauffman [KAUF93] makes an argument for autocatalytic cycles and extends the notion to other domains, deriving the notion of an *order for free* [KAUF95] that would counter the second thermodynamic law and entropy in dynamic systems [KAUF00].

**Autopoiesis**, defined by Varela *et al.* [VMU74] is the idea of a structure that is: 1. Defined in space, it has a boundary with the external environment. 2. Able to reproduce itself. This is a variation on the theme of autonomous structures and self-organisation applied in a biological setup. The notion has attracted much controversy (related by Barry McMullin in [MULL04]) as to whether it is a suitable model for living entities, and the application of the definition has itself rooted out numerous problems (such as how to define the structure boundary, and what permeability is allowed so it can interact with its environment [BEER04]). However, when viewed in a larger framework of autocatalytic cycles and autonomous structures, the concept rejoins the view [KAUF00] that self-organisation is anterior to evolution and adaptation (See also Arantza Etxeberria [ETXE04]).

**Synergetics** is the name of an inter-disciplinary approach founded by Herman Haken [HW73]. The best definition is perhaps the one given by the Center of Synergetics, headed by Haken himself: *Synergetics deals with complex systems that are composed of many individual parts (components, elements) that interact with each other and are able to produce spatial, temporal or functional structures by self-organisation*[3]. The initial topics of investigation were focused on physics, but the field has enlarged and the current domains of research of the institute are brain theory and psychology. As the etymology "science of synergy" suggests, "synergetics" calls preferentially for a holistic approach of emergence. Carlos Gershenson also proposes in [GERS07] a methodology for controlling complex systems that is well suited to this approach.

### 2.1.2 Descriptive qualifiers of emergence

The definitions presented in this section are used to classify the different kinds of complexity, emergence, or properties the entities under investigation should or should not have. However they generally do not bring any predictive power.

---

3 From http://itp1.uni-stuttgart.de/en/arbeitsgruppen/?W=5&T=1, 2008/01/15

**Nominal emergence** refers to a global property that cannot be a micro-property, like the total volume, colour, or temperature of an object. As the etymology suggests no additional assumption is imposed on the emergent notion. Nominal emergence does not refine what are the expected properties for the different levels of investigation. Thus, a nominally emergent phenomenon in a given context might not be considered emergent in another, depending on these contexts particular assumptions. To illustrate the problem let's consider the example of the colour "green", which might be associated to a range of wavelengths. But one might be interested in why the object emits these particular wavelengths (at the atomic excitation level for a LED, or through diffraction for a rainbow, etc.); or why "green" was associated to that particular range of wavelengths (which is related to the presence of receptors in human eyes, is green still green for colour-blind people?); or why we semantically associated various hues together in the same "green" concept (there might be cultural variants, so "green" is not a universally defined notion in terms of wavelengths). Nominal emergence just states the micro-macro relationship problem without hinting at the solution.

**Basic emergence** is defined by Aleš Kubík [KUBI03] as a *behavior reducible to agent-to-agent interactions without any evolutionary process involved. [...] The environment has no rules of behavior and is changed only by the actions of the agents. [...] Basic emergence then refers to a property of the system that can be produced by interactions of its agents (components) with each other and with the environment and cannot be produced by summing behaviors of individual agents in the environment*. This definition is applicable only in contexts where "agent" and "environment" have a signification, and requires that we can somehow measure the behaviour of the agents as well as define the lack of evolution. In the context of [KUBI03] grammars are used for representing the agents and their interactions. It is certainly useful to compare explicitly what are the sum and the whole, but the definition would require some adaptation to be applicable to other contexts.

**Dynamic and static emergence** as introduced by Russ Abbott [ABBO06] refer to whether a temporal aspect is respectively necessary or not for the definition of emergence. For example, diamond and graphite exhibit different statically emergent properties of carbon, like the hardness property. Dynamic emergence is **stigmergetic** when it involves autonomous entities, with an autonomous entity defined *as a self-perpetuating region of reduced entropy that is implementing a dissipative structure's abstract design* [ABBO06] (see also the corresponding entries in Section 2.1.1). In addition, a requirement is introduced that the emergent phenomenon *may be understood in its own terms* and that *its understanding does not depend on knowing how it is implemented*. This further restricts emergence to functionally irreducible cases with a formal higher-level system on these functions so they can be understood. But then, semantic closure (see below) have to be considered for how these functions and formal system relate together.

**Syntactic and semantic emergence** proposed by Howard Pattee [PATT95] are respectively concerned with the formal and functional aspects of an entity. Given a formal lower-level system, like a grammar, the syntactic emergence refers to how an entity defined at a higher-level of investigation appears in the lower level. The semantic emergence claim is that some function of the entity may not be described within the formal lower-level system. So as to illustrate the notion let's consider the dictionary example: It may be seen as a directed graph of words, each word pointing to some other words in its definition, with locally ordered graph edges. Yet, the precise meaning of a word is not contained in the dictionary itself, but found only with respect to prior knowledge at the higher level, obtained by how the language is used in practice: If each word is replaced by a sequence number corresponding to the first occurrence of that word in the dictionary the formal directed graph remains the same, but the dictionary becomes completely useless to a human.

This leads to the notion of *semantic closure*, that a higher-level of investigation is only completely defined by considering not only how the entities involved interact, but also by what meaning is associated to the interactions by an external observer or by the entities themselves. The controversy arises in both cases regarding the source of the attribution of the meaning. If it is given by the observer then it is a subjective property, not inherent to the system. Unless the observer is also part of the system, but this is equivalent to the second case that the meaning is given by some of the entities. But then, this introduces another philosophical debate as this assumes that a part of the system has the ability to attribute a "meaning" to another part of the system. Engaging in either debate is out of the scope of this document.

More generally, a functionalist approach would use semantic closure to justify the irreducibility of some higher-level function. So, being semantically emergent is possibly simply the bottom-up equivalent of being functionally irreducible in a top-down context. Section 2.2 details the notions of reductionism in relation to causality, and gives possible reasons for the controversies.

**Weak, medium and strong** emergence refer to what form of irreducibility and causal powers are attributed to the emergent entities over the lower-level from which they emerge. This is detailed in Section 2.2.

**Emergence relative to a model** does not consider emergence to be an intrinsic absolute property of a phenomenon, but that it can only be defined by considering this phenomenon with respect to an observer (which could be a formal model for example). Peter Cariani defines it as *a functional theory of emergence by giving an account of how new basic functions of the observer – measurements, computations, and controls – can come into being* [CARI89]. The observer has predictive capabilities, a formalisation of the entities and their functions at the level with which it interacts. Emergence is associated to a divergence between the model formal predictions and what really happens. The case where new observables are necessary to represent new functions in the observer model is called **creative emergence**, otherwise this is **combinatorial emergence**[4]. Note that in this context an observer is really embedded in the system under investigation, as are humans making observations about the world. Which in turn gives another view on the notion of subjectivity, with the associated philosophical controversies.

**Surprise** of the observer has been proposed by Ronald *et al.* as a condition for emergence [RSC99]. The subject is highly controversial (see [KUBI03]), mainly because of different definitions of what "surprise" means. Arguments on the subject may be classified as to whether the observer is part of the system (surprise = difference from expectation = emergence relative to the observer internal model of the rest of the system) or whether the observer is independent of the system (in which case surprise and emergence are not properly defined within the system).

**Computational emergence** is an attribute applicable to other emergence concepts. It implies the existence of a formal system, that usually allows computation theory. Any emergence definition in this context will have the "computational emergence" attribute. This tells nothing about what properties the computations and formal aspects should have to be entitled "emergent" in the first place, and what other requirements the framework must respect. In particular, this attribute alone does not specify what forms of reducibility are considered, if any. The "computational" attribute for emergence is used by people who propose that the universe is non-computable [CARI89], or that complexity is what cannot be simulated [ROSE98], so as to make the distinction with a **thermodynamic emergence** that could then be exhibited only by natural phenomena[5]. On the other side of the argument Digital Physics as proposed by Konrad Zuse [ZUSE69] and Edward Fredkin [FRED90] makes the distinction meaningless.

### 2.1.3 Quantifiable aspects of Complexity

Unlike the previous definitions and concepts, the notions in this section are not only descriptive, but do have some kind of predictive power. Hence they may form the basis for a quantitative theory of complex systems, however limited in scope this "reductionist" theory might look in a first time compared to the more elusive holistic concepts.

---

4   This short summary is far from fully rendering the works by Robert Rosen [ROSE98], Peter Cariani [CARI89], and others. I think I have captured the essence of the "emergence from a model" notion, but invite interested readers to refer to the material in [CARI89].
5   Which is a separate issue from the observer/model topic aforementioned.

**Entropy**, whether the thermodynamics or the information-theoretic version of it (Cosma R. Shalizi gives a comparative argument in [SHAL04]), has been the subject of much attention. Since entropy is associated to disorder, the idea is that organisation (and the self-version) opposes entropy and therefore we shall be able to detect it when entropy reduces. See also the entry about self-organisation in 2.1.1 and the citation from Francis Heylighen [HEYL01] in that entry. When the probabilistic definition of entropy is used, then we can actually compute it. Prokopenko *et al.* [PBR06] present an information-theoretic approach of entropy and its relation with statistical complexity. In addition, *excess entropy* is defined by James P. Crutchfield and David P. Feldman [CF03] as *the intrinsic redundancy* of the system under investigation. Together with the statistical complexity measure C (see the next entry below), the excess entropy E can be used to define Shalizi's *efficiency of prediction* indicator e = E/C [SHAL01]. In turn, this leads to a characterization of emergence: when the predictive efficiency is increased as a result of a transformation (the transformed entity can be predicted more efficiently than the original one). [SHAL01] gives an example on an ideal gas where thermodynamics emerges from the statistical mechanics.

**Statistical Complexity** measures the amount of information that is present in the past of a system, which is relevant to predicting its future. See the aforementioned [PBR06] and [SHAL06] for an introduction, [CRUT94] where James P. Crutchfield gives a link to the emergence issue, [SHAL01] for mathematics, and Frank B. Knight pioneer article on the topic [KNIG75] for the general notion on continuous systems. Statistical Complexity is the amount of information needed for optimal statistical prediction. The idea is that both well-ordered systems and highly random ones have a low complexity: The ordered systems state space usually comprises only a few states, and knowledge about these states is enough to predict the future. Random systems also require little knowledge of the past: for example, if the observed statistical distribution of events takes the form of a fixed repartition of future values, whatever the past value, then knowledge about the past is useless for predicting the future with maximal accuracy on average. Statistical complexity is thus a measure of how difficult it is to predict the future by monitoring the system past. It is defined as the amount of information present in the "causal states" of the system: the equivalence classes of system pasts that produce the same distribution of futures. Statistical complexity was proposed as a measure of self-organisation [SHAL01]: A system is said to self-organise when its statistical complexity increases over time. The measure is an intrinsic property of the system that can be computed from data [SS04]. The algorithm proposed in Shalizi *et al.* [SHRKM05] was further extended by the author of the present discussion in [BROD07] so data can now be provided on-line and the statistical complexity can be computed incrementally including for non-stationary systems.

**Algorithmic complexity** has been defined independently by Solomonoff, Kolmogorov and Chaitin (see [CHAI05] for an intuitive presentation of the concept). The idea is to use a universal Turing machine [TURI36] to describe an entity: this description then takes the form of a program. The algorithmic complexity of the entity is defined as the length of the shortest program that can produce the entity description. The problem is that this value is uncomputable and can only be approximated from above [CHAI74]. [CHAI74] also proves that *the great majority of the strings of length n are of complexity approximately n. These are the random strings of length n*. In other words, the string itself is then its shortest description, and these form the vast majority of all strings. Unfortunately, this also includes descriptions of higher-level entities, and offers no discrimination between "emergent" or "trivial" ones. In practice we may be more interested in approximate versions of a given entity and discard small variations as "noise" (see Section 2.4): When observing a phenomenon, we'd like to characterize not only the particular instances we're monitoring but also to generalise to all similar phenomena. Algorithmic complexity has a kind of continuity property, with bounds put on the complexity of entities that differ by a small variation. However Algorithmic complexity is concerned with the difficulty to describe, without a temporal aspect; Statistical complexity with the difficulty to predict, based on past instances. Algorithmic complexity gives a maximal value for completely random sequences (it seeks an exact reconstruction), Statistical complexity a minimal one (it seeks only to reconstruct a series with the same statistical properties). Depending on how we want to generalise, a definition of complexity might be better suited than the other.

**The Edge of Chaos** is an hypothetical region in parameter space between "order" on one side, and "chaos" on the other. The initial term comes from [LANG90], where Christopher G. Langton's λ parameter is hypothesised to reach a high value when a cellular automaton has a potential for complex computations. This particular λ parameter interpretation was later refuted by Mitchell *et al.* [MHC93], but nevertheless, the idea that some indicators are low for both highly ordered and highly disordered systems, while high in-between, is a very useful one: It helps characterize systems that are in a way the most "complex", these where there is the most **diversity** in significantly different global behaviours.

Indeed, both totally ordered and totally random systems lack diversity, in the sense that it is not possible to distinguish statistically the states that are produced by the system: There are only a few distinct configurations possible in ordered systems, and totally random systems exhibit the same statistical distribution of behaviours whatever the initial conditions. Therefore, the Edge of Chaos hypothesis is also that systems need to exhibit a sufficient diversity so they can support advanced features like being able to compute. This argument was proposed by [LANG90] for cellular automata, and Stephen Wolfram wrote a controversial book [WOLF02] on the notion of cellular automata exhibiting complex behaviours. Though as aforementioned the indicator proposed by [LANG90] for detecting the edge of chaos was refuted by [MHC93], the idea remains and it is possible that other indicators could work better (including for cellular automata).

Andrew Wuensche [WUEN02] extends the notions of order and chaos to random boolean networks, which are automata on a graph structure instead of a regular lattice. The large-scale dynamical properties of both cellular automata and such networks are then studied and analysed, especially under perturbation. A balance between order and chaos is then specified as a condition for the network to exhibit a form of memory. The *memory capacity* defined by Natschläger *et al.* in [NBL04] precisely quantifies with an explicit measurement the "edge" region where the system has maximum memory. [NBL04] also uses the idea that both random and ordered systems produce few indistinguishable final states, so in the context of neural networks[6] the system could be also described by its ability to separate initial configurations. This leads to the *NM-separation property*, which was presented as indicative of high processing capabilities by Robert Legenstein and Wolfgang Maass [LM07A] with an explicit mention of the edge of chaos hypothesis (see also [LM07B]).

The problem with these indicators is that they all define what [SHAL01] calls a *One-Humped Curve*, where the maximum of the curve does not necessarily corresponds to a maximum in complexity. The general Edge of Chaos claim is that some indicator related to complexity reaches a maximum between states that can be related to order and chaos, but one has yet to define what is meant by order, chaos, and complexity.

**Scale-free relations** are functionally defined by the presence of a few important elements with many less important ones, with a negative exponential relation between number and importance. The trade-off between importance (functional role) and number allows to scale the system by making it manageable as its size grows. Réka Albert and Albert-László Barabási [AB02] exhibit such scale-free relations in the domain of network graphs where a few nodes (ex: internet routers) allow efficient network traversal, by aggregating the many local sub-networks hierarchically.

A functional presentation was deliberately used here because there is a controversy as to the exact mathematical relation and the meaning associated to the exponential decrease. The historical perspective provided by Michael Mitzenmacher [MITZ04] and Edoardo Milotti [MILO01] shows that power-laws and exponential relations are widely applied and ancient concepts[7]. A scale free relation is when the probability of finding the property of interest decreases according to some power-law or log-normal distribution as the scale increases. Such relations may be found for network graphs as aforementioned in [AB02] and in [WUEN02], but they also appear in finance, biology, chemistry, ecology, astronomy, and information theory (see [MITZ04]). When taken in the frequency domain, the *1/f noise* equivalent of the power-law appears in particular in electronics, sandpile models, and more inter-disciplinary fields (see [MILO01]). Such a wide range of applications makes them a good candidate for detecting cross-disciplinary universal phenomena, hence makes them a primary target for Complexity Theory. The risk is of course over-generalisation with no predictive power, and [MILO01] concludes by: *Do we understand 1/f noise? My impression is that there is no real mystery behind 1/f noise, that there is no real universality and that in most cases the observed 1/f noises have been explained by beautiful and mostly ad hoc models.*

Yet this is precisely what a functionalist point of view of emergence would appreciate: Irrespectively of the underlying elements, the functional property of being (relatively) insensitive to scaling remains, at least over the range of scales that matters for a specific problem, and whatever the mathematical model that is best suited for the description (power-law or log-normal). For example, efficient traversal through hubs in a network is a property that is interesting both locally (with a small group of nodes connected to a local hub) or globally (for reaching distant sites). Whether the distribution of connections below each node follows a specific mathematical form matters in this case much less than the property itself.

On the other hand, in the contexts where the power law is used to make predictions in the range corresponding to the tail of the distribution, an error in the formula can have drastic consequences. So the reductionist/functionalist debate strikes again, and [MITZ04] warns: *From a more pragmatic point of view, it might be reasonable to use whichever distribution makes it easier to obtain results. This runs the risk of being inaccurate; perhaps in some cases, the fact that power law distributions can have infinite mean and variance are salient features, and therefore substituting a log-normal distribution loses this important characteristic. Also, if one is attempting to predict future behavior based on current data, misrepresenting the tail of the distribution could have severe consequences.* More generally the study of the tails of probability distributions and their decrease rate is the topic of the theory of large deviations, for which Srinivasa Varadhan received the Abel prize in 2007. This is an important mathematical topic with consequences anywhere a predictive methodology is sought in the aforementioned disciplines exhibiting "power laws" (or related).

---

6 The "neural" networks are in the [NBL04] case actually networks of transfer functions, so they generalise the boolean networks for which the functions act on and produce boolean values.

7 And so is the mathematical controversy, ex: between Simon and Mandelbrot as explained in [MITZ04]

Finally, even a functionalist might be interested in the analysis of the differences between the mathematical forms. [AB02] proposes for example a generative model for the power-laws observed in networks, but perhaps other models are more statistically significant. The reason why a property is observed with a negative exponential-like relation in a given system, what led to this relation, might be interesting in order to better understand the function the property occupies in the system and its limitations. In particular [MITZ04] concludes *The fact that power law distributions arise for multiplicative models once the observation time is random or a lower boundary is put into effect, however, may suggest that power laws are more robust models.* Thus, conversely, analysing the form best suited to model a functional property behaviour might actually be indicative of the reasons why this property occurs. And as is explained in the next section "the reason why" something happens is a causality issue, which is precisely indicative of the functionalist/reductionist debate.

## 2.2 The tricky concept of causality

The concepts presented in the previous sections give an image of a fragmented field, where controversies abound. This is precisely the case, and current attempts at creating a theory of emergent phenomena often end up having to define concepts that are specific to that attempt. Jaegwon Kim [KIM06] notes that *Emergence is very much a term of philosophical trade; it can pretty much mean whatever you want it to mean, the only condition being that you had better be reasonably clear about what you mean, and that your concept turns out to be something interesting and theoretically useful.* Consequently, there are as many definitions as frameworks, and no real common theory.

Yet, many if not all emergence-related concepts in the previous section refer to some form of (or lack of) causation. As mentioned in introduction this is expected since "emergence" is precisely a term which is invoked when other explanations fail. Hence causation is the subject of controversies: If no reason can be given for "emergent" behaviours, why do they appear? The debate between reductionists and functionalists related by Jaegwon Kim [KIM99] revolves around the same idea. Jochen Fromm [FROM05] notes: *the cause is normally the unclear point in emergence* and proposes a taxonomy of emergence concepts based according to how they treat causality. Causality is also analysed as a major source for historical debate by Peter A. Corning [CORN02], where finality (functional causality) is explained in detail in the section about synergism.

Hence, solving the problem of "why" certain phenomena appear in some context may thus very well put these phenomena outside the category of "emergence". Of course this definition of "emergence" is circular, hence not really a definition: emergence would be when micro-macro relations are too complex to be understood, and complex systems science the study of emergence. Unfortunately many of the concepts presented in the previous section also rely on such circular definitions. For example, self-organisation is often given a causal power, while not even defined unambiguously. By examining the notion of causality with respect to its relations with the emergence issue, this section goal is thus to analyse what are possible reasons for usual controversies and to propose a disambiguation.

### 2.2.1 Causality as a source of debate

In his review about the history and *re-discovering of emergence*, Peter A. Corning [CORN02] traces back a major source of controversy to the notion of causality. The two sides of the argument are presented as the "holists" on one side, and the "reductionists" on the other, with radically different perspectives on causality which are detailed below.

Mark Bedau in [BEDA03] states that *emergent properties without causal powers would be mere epiphenomena*. Russ Abbott in [ABBO06] states: *In short, we define epiphenomenal and emergent to be synonyms*, but then he puts the debate between "reductionists" and "functionalists", the later ones being equivalent to the aforementioned "holists".

In order to clarify the concept, this document reuses the classification that Emmeche *et al.* have outlined in [EKS00] by applying to the question of emergence the four Aristotelian concepts of causation:
- **Efficient causality** is the notion that something implies, entails, or brings about something else.
- **Material causality** is the notion that something is made of something else. Note that "matter", as in material, has the broader sense of "composition" here.
- **Formal causality** is the structure or the form of something, like a house is defined by its architecture.
- **Functional causality**, which replaces finality in Aristotelian terms: the role played by something (in relation to something else).

For example, discussions about an alarm clock may refer to the formal causality (the clock internal plan, why it works), functional causality (what the clock is used for, why it was built), material causality (the clock composition, why it exists at all), and efficient causality (the clock is the cause of the sound that is itself the cause of the observer waking up).

So, what are the two sides or the argument? Holists/functionalists are more concerned with the functional causality, whereas reductionists are more concerned with the material and formal causality. The clash often comes when the two camps refer to their favourite concept to explain something, thus bringing efficient causality in the balance. On the one side, the function of something is the ultimate source of why things happen, and on the other the explication comes from material and formal laws of operation.

Many apparent controversies end as soon as the notions of causality are refined. For example: "The whirlpool causes the water molecules to move in a restricted way" versus "Water molecules and heat processes amongst other things, are the cause of what we perceive and define as a whirlpool". In this case, the first statement would be an efficient causality (the restriction) between objects defined functionally and formally (the whirlpool and the water molecule movements). The second statement is about an object defined materially (the whirlpool). Natural language only is the source for a possible confusion: Applying the efficient causality of the first statement to the whirlpool of the second statement is meaningless and should be discarded as such (from the second point of view, the restrictions are part of the definition, not a consequence).

Is that all? Can causality and all controversies be solved by referring to this simple classification? Of course not, but it makes a good start. Further refinements could be made using notions like time dependency, what is required for objects to be comparable, probabilities, and more. Howard Pattee [PATT97] proposes that causation is a useful concept only when it identifies controllable events or actions. This is further extended by Fabio Boschetti and Randall Gray in [FG07], who propose a form of causation intermediate to the above four, as exemplified by: *The flock will circumvent the obstacle. It thus appears that we were able to exert control on the behaviour of the flock; the flock appears to have causal power.* The causation question then becomes identifying what "controllable" means, with the related philosophical issues that are out of scope of this document.

Modern physics must deal with the quantum principle of no local reality and Bell's inequality violation, combined to the no communication principle, so as to avoid a time travel paradox in general relativity. The no local reality is in apparent contradiction with material causality, the no communication principle restricts efficient causality. Measurements may become important, since they can provide an objective source of investigation for material causality. But unfortunately, as mentioned by Howard Pattee in [PATT95], a measurement is only defined by the function of the measuring device: to provide a number, that is interpreted in the light of a theory. The theory then itself provides formal causality between the measurements, by way of its laws. As we see, the problem of causality is intrinsically linked to the problem of material objectivity.

### 2.2.2 Supervenience and identity

Supervenience is typically used to assume material causality while avoiding the issues related to other forms of causality. For example, saying that the mind supervenes on the body means that ultimately the body is the material source of the mind, without assuming anything as to how the mind may "emerge" from the body.

More precisely **supervenience** is concerned with a logical dependence between properties. Assuming properties A and B are defined, A supervenes on B means that each time entities differ with respect to property A, they also differ in property B. This means that no two entities may have the same B without having the same A. The difference is purely theoretical: whether we have the means of investigating this difference or not is out of topic for supervenience. The supervenience concept is also not concerned with "levels", just properties. These properties may be defined, or not, at different levels of investigation. A stronger version has also been proposed[8]: A property A strongly supervenes on a property B whenever each time it is possible to define properties A and B in a framework, no entity could differ in property A without also differing in property B, whatever the framework.

Supervenience represents a weak form of micro-to-macro relationship that still has useful consequences. Assuming supervenience allows to reason at a high level (for example on the movements of billiard balls) and then to apply the result of this reasoning on the micro level through material causality (for example deducing that atoms have moved, even though the collision laws that apply to billiard balls do not apply directly to the atoms). This is what Russ Abbott calls *downward entailment* in [ABBO06], and which is outlined with more details in Section 2.2.4.

But as mentioned in the previous section, controversies occur when mixing different notions of causality. As an example let's consider a diamond made of carbon atoms. One could say that the diamond supervenes on the carbon atoms: when considering a particular, unique, set of atoms, one must also consider a particular, unique, diamond. No two diamonds may be made of the same atoms: this is material causality. On the other hand, if talking about formal causality, a diamond is generically "made of" a pattern of carbon atoms, and the atoms are all alike so we don't really care which specific atoms are used. A diamond is a specific pattern in carbon atom organisation, which distinguishes it from graphite: both could be defined with the exact same atoms, but their organisation is what matters. Therefore, the diamond also supervenes on the carbon atom organisation: two measurably different diamonds will have a different pattern, two exactly similar diamond, down to the atomic level, will have the same pattern.

But what does it mean to be "the same"? Equivalently, for the purpose of the supervenience definition, what does it mean to be "different"? Is a reproduction "the same" as the original? The philosophical controversy arises when one chooses a different form of causality for the notion of identity, like the material and formal examples above. Digital objects are more concerned with the formal aspect, famous paintings with the material one, but what about the material reprint of an original digital artwork uniquely displayed for a specific exhibit? Then, there is

---

8 For a more complete discussion on the various forms of supervenience see for example the Stanford Encyclopedia of Philosophy at http://plato.stanford.edu/entries/supervenience (2008/01/15) also a functional (social) dimension to take into account. In some cases the material or formal identity does not matter as much as the functional identity. For example, when using a boat to escape a flood it doesn't matter whether the boat is made of wood or tin, or what form it has, so long as it floats. In this example "boat" has the functional identity "something that floats", irrespectively of the material or formal identity.

Compared to the weaker version that has just been explained, the stronger version of supervenience implicitly assumes we can define "the same" properties A and B across different frameworks. Of course, depending on the chosen perspectives for defining sameness in entities and sameness in properties, this stronger version may range from a tautology to a puzzling issue.

Without engaging in the controversy, that is assuming a particular definition for "sameness" has been given for a context, supervenience can then be used. However, in order to prove (or disprove) supervenience, one would need to derive an investigation tool that can precisely identify differences in the chosen properties. In the case of emergence between two levels of investigation, proving supervenience in practice would require measuring the exact state of the lower-level system. This is assuming such measurements do not themselves modify the lower-level state, as is the case in quantum physics. Without such a tool, the only remaining possibilities are to accept or reject supervenience as an axiomatic property of the system, or to build explicitly a system in which it holds. That last explicit system building scenario includes the case for deterministic computer simulations, so supervenience holds by definition for the examples in Section 2.4.

In any case supervenience does not help much for understanding the micro-macro relationship. The Wikipedia entry about supervenience[9] notes: *Supervenience has traditionally been used to describe relationships between sets of properties in a manner which does not imply a strong reductive relationship. [...] Supervenience allows one to hold that "high-level phenonema" (like those of economics, psychology, or aesthetics) depend, ultimately, on physics, without assuming that one can study those high-level phenomena using means appropriate to physics*. The next section deals with the micro-macro relationship, the problem of finding what is the cause of a given high level phenomenon. Section 2.2.4 deals with the strength of the reductive relationship that is mentioned in the Wikipedia citation.

### 2.2.3 Causal reductionism

This section deals with one of the major controversies: whether and perhaps more importantly how an "emergent" phenomenon is reducible or not to the lower-level elements and interactions from which it emerges. The different notions of causality that were previously introduced are analysed with respect to their relation to reductionism. This section thus deals with the bottom-up causal link. The next section deals with the other major controversy, related to the top-down causation.

Causal reductionism is the assumption that every phenomenon, whatever its level of investigation, ultimately have a cause, except possibly axiomatic properties which are postulated. If additionally a unique cause is assumed to have a unique effect then supervenience holds.

Depending of the causality perspective chosen, causal reductionism has different consequences. Material causal reductionism states in essence that whatever observed complex phenomena, they are always made of matter (in the broad compositional sense), be it an electron stream inside a computer or a magnetic field around the galaxy. Of course, material reductionists do not reject phenomena like consciousness or social constructs like flash mobs. It's just that stating that a brain and a crowd are made of atoms does not help much in understanding these phenomena, hence material reductionism may not be the best notion to use in these cases.

Formal causality gets around the problem by stating that brains and crowds additionally have an internal structure and governing laws that must be considered. Formal reductionism is then the assumption that such laws can always be found, that any higher-level effect is logically connected to the lower-level formal system. Unfortunately, most formal systems are known to be incomplete: No amount of formal causality may satisfyingly encompass all higher-level constructs. If, by analogy with [CHAI05], such intrinsically logically undecidable higher-level phenomena are the vast majority of all higher-level statements[10], then the question becomes whether these statements are really observable or not. Of course, it is still possible to postulate (whether this is true or not) that reality and all higher-level measurements are logically reducible. Therefore by definition any observer, part of that system, whatever its level of investigation, can only observe logically reducible statements. But even then, the observer may not be able to take advantage of the reduction in any efficient way: this would assume the observer has total knowledge of the underlying rules (which does not generally hold) and that it seeks perfect reconstruction, even for computationally incompressible statements (which is usually not what we want to do, see Section 2.4). Assuming formal reductionism or not is a matter of principles, and doesn't help much for practical investigations[11].

---

9 http://en.wikipedia.org/wiki/Supervenience, version 07:11, 17 November 2007

10 Gregory Chaitin [CHAI05] gives a special attention to the case for real numbers in particular and to the limits of formal systems in general. The statement in the main text is not a citation, just a reformulation of what I think is a main idea in [CHAI05].

11 Unless one already has a working formula, in which case, of course, this paragraph doesn't apply. However taking advantage of a formula to relate the lower formal level to an observed higher level

Functional causality offers another relation to reductionism. In this setup, higher-level phenomena are defined by their relation (function) with other higher-level phenomena and the environment at that level. Functional reductionism is a contradiction, since the function is by definition only a higher-level construct. This is also the main holist approach: assume the irreducibility of the function to lower levels and consider only functional causes at the higher level. In that case, reductionism would take the form of assuming every phenomena has a function. But how is this function defined? It is not possible to isolate one part of the system and assign it a function independently of the rest of the system: Dependency loops are inherent to functional causality. An elaborated view on self-reference and an introduction to Howard Pattee's semantic closure concept can be found in [PATT95]. There may be dependency chains, which can be given some degree of functional causal power, but as soon as a loop is reached, the reduction argument breaks apart. For that reason, functional causality alone cannot meaningfully be associated with reductionism: there may be reductionism in a system, but then functional causality will not be the only causal relationship in that system. Functional irreducibility is thus only meaningful as a definition, that a function is only defined at a high level.

The fourth form of causality mentioned in Section 2.2.1 is the efficient one. Unfortunately, pure efficient causality also suffers from infinite regression. When given an efficient causality chain, one can always backtrack to the proximal cause, without end, so long as one stays purely in efficient causality. To break the chain, one requires another form of causality (such as material of formal). But then, the argument falls back to one of the previous points. Aristotle broke the argument at the other end of the chain, by stating the entities act according to their finality or purpose, which was relabelled the functional cause in Section 2.2.1. Once again, we're back to another form of causality. Reductionism is not a meaningful concept in pure efficient causality terms.

Are there other forms of causality one could apply reductionism to? Perhaps, but as previously discussed, these would certainly also come with their own lot of limitations. For example reductionism applied to the aforementioned notion of control also suffers from infinite regression (controlling the way to control the way... to control an effect). Amongst the four forms of causation presented in Section 2.2.1 the only one that is consistent with reductionism is the material one, and possibly the formal one too by construction or postulate. But as noticed, neither one helps much in understanding the emergence issue in general: emergent concepts are usually associated with a high-level functional definition.

---

phenomenon would put that phenomenon outside the scope of some emergence definitions presented in Section 2.1.

### 2.2.4 Downward causation and the strength of emergence

Downward causation is the statement that some higher-level construct may exert causal power on the lower-level. This is the inverse problem as the one detailed in the previous section. The controversies are once again associated to what exactly one means by a causality relationship, as reviewed by Jaegwon Kim [KIM99]. An illustration of downward causation with efficient consequences would be the placebo effect, if this effect really exists[12]: when a patient is given sugar pills instead of active drugs and still reacts as if she/he would have received a real medicine. In this case the mind would have a downward effect on the body.

Emmeche *et al.* [EKS00] distinguish between three types of downward causation: strong, medium, and weak. **Strong downward causation** is the mix of constitutive irreducibility and substantial realism of the higher level. The medium version is the combination of constitutive irreducibility, formal realism, and a refinement detailed below. The weak version is constitutive reductionism, formal realism, and a stronger version of the refinement.

Constitutive irreducibility is another way of saying that material reductionism alone is not enough: The building blocks that make up the higher level are assumed to involve a materially irreducible part. Substantial realism additionally claims that these new building blocks are matter as such, in the broad sense of a part of reality, that is "matter" at the higher levels is as valid as as "matter" at the lower levels. This amounts to the creation of new fundamental matter (broad sense) *ex nihilo*, and two identical low-level states could lead to distinct high-level ones thanks to the presence of new compositional matter at the high level. This contradicts supervenience of the emergent property on the low levels, as noted by Jaegwon Kim [KIM06]. When considering downward causation, the strong emergence requirements additionally state that new entities have material causal powers downward. Emmeche *et al.* [EKS00] give the example that strong emergence is like considering that *the emergence of the cell as a living substance efficiently causes changes in the molecules, making them somehow specifically "biological"*. This form of emergence is usually only defined so as to be rejected, and Jaegwon Kim [KIM06] asks *whether it is a form of emergence at all.*

**Medium downward causation** replaces the strong substantial realism requirement by a formal realism one, and adds another requirement detailed below. Formal realism does not mean formal reducibility, it means that unlike the previous case, the material component of the higher level may be formal elements of the lower one as opposed to material ones. Both requirements are generally accepted so long as these new entities are only considered from that level up, and so long as one defines "reality" from the higher-level entities point of view. For example, a simulation could be "real" for the agents in it, the "matter" the agents manipulate is real for them but only formally defined for the lower-level. The problem comes with the downward causation argument. At this point, no restriction has yet been put on what the agents are allowed to do on the lower level. The 'biological molecules' example given above for the strong version of downward causation could still have its formal counterpart with laws that are specific to the agents. For the medium version of the downward causation concept to be viable there must be some limitation that prevents the higher-level irreducible entities to modify, restrict, or more generally change in any way, the lower-level formal rules that lead to their existence. To use another example, so far, the mind could alter physical laws.

---

12 The placebo effect is quite controversial, as exemplified by the heated argument between Asbjørn Hróbjartsson and Peter C. Gøtzsche [HG07] on one side, and Wampold *et al.* [WIM07] on the other. Pain treatment seems to be the domain with the least controversy, though even in this case the existence of a placebo effect is statistically hard to assert. In general the placebo effect, if any, strongly depends on the experimental conditions, as well as on what symptoms are treated.

The term **strong emergence** usually[13] refers either to (or both):

- The first creation of new compositional matter out of nothing.
- The lesser form of creation or modification of lower-level laws or effects.

Emmeche *et al.* [EKS00] refine the concept for medium downward causation by adding the requirement of an efficient causality restriction, which includes a temporal restriction. This requirement excludes from medium downward causation any change in the formal laws, together with any back-in-time change in initial conditions. The agents can no longer modify lower system laws in any efficient way, and the mind can no longer change the physical laws. In other words, the higher level entities can only constrain the domain of future possibilities of the system compared to past history, which is reminiscent of the cognitive domain notion as defined by Randall D. Beer [BEER04]. However, this does not preclude a unique lower-level state to coexist with several different higher-level entities, what Emmeche *et al.* [EKS00] call inverse supervenience.

One could argue that according to formal irreducibility, some phenomena are not logically reducible to lower-level rules and may thus be accepted, or not, with the same lower-level state. In the same way, it is possible to subscribe to the axiom of choice, or not, in ensemble theory. But then, such a phenomena cannot by definition have any downward causal power, which contradicts the downward causation concept.

Another interpretation is given by Emmeche *et al.* [EKS00]. By analogy with dynamical systems, the concept of boundary conditions is introduced. Medium downward causation would take the form of an influence of the higher-level concepts on the shape of the phase space, by changing some parameters, or by restricting the boundary to some region. Nevertheless, Emmeche *et al.* [EKS00] still do not fully clarify this inverse supervenience concept: their conclusion ends up with applying dynamical system rules only *in a somewhat metaphorical sense.*

An additional potential issue with the insertion of that restriction for medium downward causation, is that it excludes some phenomena like the placebo effect[14]. This effect can be seen as a downward efficient causation from the mind on the body. Of course, the downward aspect depends on the perspective chosen for what is the mind, especially what kind of reductionism is assumed or not. In any case, Emmeche *et al.* [EKS00] do not pretend to solve all the controversies associated to downward causation; they propose an interpretation framework that admittedly does not cover all cases.

Jaegwon Kim [KIM99] proposes that the downward causation concept should be replaced by a **downward causal explanation** one: whether the explanation is given in terms of higher or lower level concepts. Jaegwon Kim [KIM99] concludes that while this may not be enough to *save real downward causation, perhaps that is all we need or should care about.*

The **weak downward causation** version is not affected by the inverse supervenience problem. As in the previous examples, there are new irreducible higher-level constituents due to formal realism (the question of what form of irreducibility is explained below). Formal realism also precludes new material effects to appear from these entities at lower-level, as was the case in the strong version. But unlike the previous examples, the constitution or composition of these new higher level entities is assumed fully materially reducible: the matter of the higher level is made of lower-level matter. This eludes the problem of the inverse supervenience of the medium downward causation case: material supervenience holds for weak downward causation. As for the medium case, an additional requirement states that weak downward causation cannot be interpreted as any kind of efficient causation.

So, what can be the non-efficient downward causal power of a fully materially reducible effect on the lower-level? Emmeche *et al.* [EKS00] give an example in terms of attractors of a dynamical system. If some higher-level concept is identified with being in an attractor basin, arguably functionally irreducible, then the downward causation is associated to the fact that the lower-level variables can only take some values in that basin and not others. The higher-level notion has "restricted" the lower-level capacities, though in this case Jaegwon Kim [KIM99] would rather say this is just a downward causal explanation, as the restriction is inherent to the system. What's not clear in the argument by Emmeche *et al.* [EKS00] with respect to weak downward causation is the type of irreducibility it allows. Given that material reducibility is assumed by definition, and given the remarks of Section 2.2.3, we may assume that only a computational incompressibility and functional irreducibility is possible with weak downward causation as defined by Emmeche *et al.* [EKS00] (not a formal one). Then, downward causation takes the form of a restriction on the lower-level possibilities. The question is then the extent of this restrictive power.

---

13 But not always clearly. Mark Bedau [BEDA03] for example defines strong emergence when emergent properties are supervenient with irreducible causal powers. This formulation is confusing as it does not specify which causal powers and which forms of supervenience to consider. Some combinations are contradictory, but some others are valid, like the downward entailment definition given further on in the main text. Jaegwon Kim precisely states that supervenience and irreducibility are two necessary but not sufficient conditions for emergence [KIM06], and notes that *how reducibility is to be understood in this context will require some discussion*.

14 Though that may not be a problem for this particular example if the placebo effect is non-existent.

**Weak emergence** is defined by Mark Bedau [BEDA03] when a higher-level property is underivable except by a full simulation (no shortcut can be found). This framework assumes material and local formal reducibility: The higher-level phenomenon under consideration for weak emergence must be fully reducible to a set of micro-effects that is "local". Nothing is said or even implied for other macro-effects using micro-effects outside this local set (the system is perhaps not globally formally reducible). Weak emergence is then equivalent to computational incompressibility (see Gregory Chaitin [CHAI74]) over that local set. Weak emergence rules out the medium and strong versions of downward causation.

Russ Abbott defines in [ABBO06] another concept related to downward causation: **Downward entailment**. Downward entailment is an effect that is defined in a framework "functionally irreducible" together with "materially and formally supervenient". Unlike the previous weak downward causation concept by Emmeche *et al.* [EKS00] the introduction of a requirement about no efficient downward causality is not needed any more thanks to supervenience, as is detailed below.

The combination functionally irreducible / formally supervenient is the one that makes the explanations in [ABBO06] confusing at times. However, there is no contradiction. Russ Abbott takes as an example the Turing machine implementation using the Game of Life. The function performed by a Turing machine is not logically deducible from the game of life rules alone: this requires higher-level concepts, the program and the machine itself. When considered solely as a precise arrangement of game of life cells these concepts make no sense. In other words, the function is irreducible, but the formal aspect is supervenient (different Turing machine states necessarily imply different cell configurations). The formal aspect of Turing machines and all computability theory is not reducible to the formal rules of the Game of Life. But the reason is the functional irreducibility that comes in between, otherwise there would be no reason why formal higher-level abstractions should be considered independently of the formal lower-level ones.

Once this issue is clarified, downward entailment amounts to reasoning formally on the higher-level to infer lower-level properties, using negative logic and supervenience. In other words, thanks to the supervenience part of the definition, it is possible to reason on the functional part. What this means *is that billiard balls, gliders, Turing Machines, and their interactions can be defined in the abstract. We can reason about them as abstractions, and then through downward entailment we can apply the results of that reasoning to any implementation of those abstractions* (Russ Abbott, [ABBO06]). The first part is a functional interpretation (billiard ball), with its associated formal system (reasoning about). Then, thanks to supervenience (any implementation), the results of the higher-level formalism may be propagated to the lower-level.

In this example, using Newtonian physics on the billiard ball will put constraints on its lower-level material and formal implementation. This is saying that since no two higher-level balls may have the same lower-level implementation (the same atoms), results about a higher-level ball must necessarily involve its unique implementation. As in the game of life example aforementioned, the higher and lower levels formal systems are disconnected: The Newtonian laws alone do not apply to individual atoms directly, other effects must be considered (the ball internal cohesion, etc.). Downward entailment, by assuming supervenience, is a way to reconnect the formal systems after they were disconnected by the functional irreducibility.

There are undoubtedly many other possible variations on the subject of downward causation, and some are given by Emmeche *et al.* in [EKS00]. So long as the hypothesis are well-defined, the academic issue is then: Can we test, validate or refute these variations distinctive properties, or are they purely theoretical?

## 2.3 *Formally irreducible emergence*

Both Mark Bedeau [BEDA03] and Russ Abbott [ABBO06] insist on the fact there is no intermediate concept between strong emergence and causal reductionism. As was explained by the previous sections such a statement requires clarification as to what form of reductionism is considered. In these [BEDA03] and [ABBO06] cases, material and formal reductionisms are assumed (deriving from a simulation assumes formal reductionism, even if only locally), but not a functional one (that's the whole point of weak emergence). This section presents the case for "functionally and formally irreducible" together with "materially reducible". This precisely forms an intermediate concept, though as we'll see further on, not a particularly useful one in practice. However, by analogy with the presentation by Gregory Chaitin [CHAI05] such concept should in fact be the predominant possibility.

The problem is related in part to the incompleteness of formal systems, which is discussed in this section, and in part to what we really want to do with these formal systems, which is detailed in Section 2.4. Given a sufficiently complex underlying micro-level system, there exist macro-level statements which are not provable[15] (whether positively or negatively) using only this system micro-level framework. The question of why and when such formally unprovable statements are observed in practice is addressed in the next section.

These statements are stronger than Mark Bedeau's weak emergence [BEDA03], in the sense that any simulation of the macro-level effects would represent a logical "proof", hence these macro-level properties are not weakly emergent. Of course, material reducibility may very well still hold, depending on the physical definition chosen for "material". But our formal equations cannot explain all higher-level observations. This irreducibility problem is generic, fundamental, and cannot be ignored.

The above irreducible statements could at first glance seem to be related to strong emergence. However, they have some crucial properties:

1. Their only consequences are necessarily expressed in "higher-level" terms, whatever that means in a particular context. By definition, if such an irreducible phenomenon could have consequences on the level at which the corresponding statement is defined, then this would negate the unprovability. For example the halting problem does not have consequences on the automaton rules themselves. The only consequences in that case are on the higher-level of the "program" and its execution in time. Of course the boundary between levels may sometimes be unclear as commented in Section 2.1.1, and occasionally the original formal system may be expanded to new axioms. But then we're really considering another, different, system with it's own higher-level unprovable effects. As Russ Abbott points out [ABBO06], at the lower-level the fundamental forces and particles of physics are already irreducible phenomena we use as axioms for the lower level realism and formalism.

2. There is no practical way to distinguish between a logically irreducible effect at a higher level, and a logically reducible but computationally incompressible one. By analogy with the demonstration by Gregory Chaitin [CHAI74] or with the seminal article by Alan Turing [TURI36], the problem of identifying a particular phenomenon as logically reducible or not is itself undecidable. As a proof sketch, let's consider that one could define an order for the different possible simulations by size, for example using the same binary coding as in [CHAI74], or the *enumeration of computable sequences* from [TURI36]. Then let's try all simulations one by one in order. If we find a simulation that produces[16] the phenomenon, fine, we've proved it is both logically reducible and computationally incompressible (we found the shorter version). Otherwise, there is no way to decide when to stop, there is the possibility a larger simulation produces the desired phenomenon: We can't decide on logical reducibility. Consequently, **given a functionally defined higher-level phenomenon, there is no general way to distinguish whether it is formally reducible but incompressible or formally irreducible**.

3. Given the difficulty to "revert" even simple deterministic chaotic dynamical systems to their initial conditions and evolution rules, exhibiting a logical reducibility for a given practical problem (and not a suitably designed scenario) may be computationally very complex. Not only is it impossible to distinguish between a theoretically logically reducible or not higher-level phenomena in general because this would be undecidable, but proving reducibility for the systems that are theoretically reducible is probably intractable on any real-sized problem.

What about strong emergence? Either a phenomenon is logically reducible to micro-effects, in which case it is a case of formal reductionism, not a case of strong emergence. Or it is logically irreducible, but then, the first point above is in essence a rejection of the strong downward causation hypothesis in that case. Therefore, the combination of both logical irreducibility and strong downward causation is a contradiction: this rules out strong emergence.

What remains are logically irreducible phenomena that do not have any effect at micro-level, though they may still be reconnected to the lower-level by using supervenience as previously mentioned, which provides a form of downward causal explanation or entailment as Russ Abbott [ABBO06] puts it. But according to the second property above, these logically irreducible phenomena are not distinguishable in practice from weakly emergent ones: The problem of deciding whether a particular statement is logically irreducible, or logically reducible but computationally incompressible, is both theoretically undecidable in the general case and probably practically intractable for the exceptions anyway. So, this explains why previous works using the weak emergence concept still remain valid: Even if irreducible phenomena (logical or incompressible) would be much more frequent than reducible ones by analogy with [CHAI05] so we probably have already met some, we can't make the distinction in practice. And in particular for what Russ

---

15 In this section we're concerned with the limitations on weak emergence as defined by algorithmic incompressibility, which is precisely the framework in which Alan M. Turing [TURI36] notion of uncomputability has consequent implications as was demonstrated by Gregory Chaitin [CHAI74]. The relation with Gödel's theorem is provided in Section 11 of [TURI36].

16 The difference between the present case and [CHAI74] and [TURI36] is that the phenomenon under investigation is defined functionally at the higher-level, not formally from within the lower-level system. Provided we have a way to test whether the phenomenon is equal or not to a simulation result, then the suggested proof is essentially the same as in [CHAI74] and [TURI36].

Abbott calls a *very complex autonomous, self-sustained entity, whose functional definition is linked to other such autonomous higher-level entities and their environment* [ABBO06].

Pure formal reductionism for all higher-level entities is insufficient in the general case due to incompleteness. Strong emergence was rejected. As mentioned in the introduction to this section, the only remaining possibility is the one that was dismissed by both Russ Abbott and Mark Bedau: an intermediate level between formal reductionism and strong emergence. Since it is undecidable whether an observed functionally defined entity could be formally reducible or not, that intermediate level both complements and is indistinguishable from weak emergence. Let's call it **formally irreducible emergence**, for lack of a better term.

One may then legitimately ask whether formally irreducible emergence is at all observable in a computer simulation. The next section investigates why it is in fact observable, and why the formally reducible or not aspect of an observed entity is actually unimportant.

## 2.4 Implications for formal systems

### 2.4.1 Analysing the results of simulations

Weak emergence is not a very useful concept for "understanding" an emergent phenomenon in practice. Of course, assuming we could obtain a simulation equivalent to running the system itself, then possibly we could make predictions if that simulation can be made to run faster than real-time. This is certainly useful, and to a certain extent this is how we already use numerical simulations, especially in industrial contexts.

However, the full simulation tells nothing about understanding the higher-level phenomenon as such. Understanding involves abstracting notions and entities that we can relate together by reasonably concise statements (compared to the simulation) that still produce good approximations. Consider as an analogy saying that as long as one sticks to the exact wave functions of quantum physics then quantum physics apply and the object is reducible to waves. But in the case of considering a macro-level object (like a stone, a flower...) what we want is usually not to consider it as an intractable bunch of waves, but rather to find how it relates to other objects at its own level.

The task is thus to find simple relationships that describe entities and their interactions with good accuracy, so our limited human minds may comprehend them. Steven Weinberg [WEIN02] says that Science is concerned with simple things. A corollary is the negation of the possibility for an objective definition of emergence that corresponds to our intuition. An advanced futuristic artificial intelligence or an hypothetical alien entity could very well label as trivial phenomena our brain structures have no chance to comprehend.

Let's now consider a computer simulation which can be made fully deterministic and reproducible. If one sticks to the exact observations obtained from that simulation they are surely formally reducible (though perhaps incompressible). But what we want is to find reasonably concise and precise approximations of the higher-level entities and their behaviours that are produced by the simulation. Does formal reducibility still holds in this case? Can the simulation produce observable and reproducible phenomena, functionally defined such that they are irreducible to the simulation program laws?

### 2.4.2 Examples

Let's consider for the purpose of this argument that the observer has total knowledge of the underlying rules, which does not generally hold if the observer is part of the system, but which is reasonable in the case of a programmer examining a computer simulation. Each observed statement is then perfectly logically reducible, though some statements are computationally incompressible (no shorter simulation can be found). If what we want to do is finding a shortcut, a concise and reasonably precise law that can describe the observation, then there is no guarantee that the approximation is itself formally reducible.

#### Example 1: Generalisation across simulation runs

Let's write a program that plots the Riemann Zeta function on the complex plane strip with real part between 0 and 1. We then observe that zeroes appear exactly on a straight line for all the simulation runs, and for as precise a result as we wish by setting the floating-point resolution. Can we generalise to all future runs?

#### Example 2: Generalisation by formalising higher-level laws

We are given a complex simulation in which some results always appear nearby a simple curve (parabola, line, exponential...), but there are small variations. As is usual in physics, let's consider these variations are noise and then derive a law with the curve to predict the coming up of new points with good accuracy[17]. The curve cannot then be directly related to the lower-level system: the formal reduction applies only to the exact points that were produced, including what was considered noise. And even without noise what was really obtained is only just a mathematical conjecture, as in Example 1, and there is no guarantee one could formally generalise to other simulation runs. But the higher-level shortcut is

---

17 Many physical laws work this way: we build descriptive laws of motion, heat propagation, etc... that give a reasonable approximation of the corresponding high-level effects. Then part of what we call "noise" includes the variations against these imperfect approximations.

Page 15/24

potentially useful, an "emergent" law from the simulation. We may consider the emergent law on its own level as an entity in itself, and use it there as we would for "physical" laws. One could then try to apply the scientific method on the higher-level as suggested by Jochen Fromm [FROM06], by defining experiments to get the limits of that law, check conditions whether it applies or not, etc. Without caring one way or another for the reducible to the lower-level or not aspect of the higher-level entities.

*Example 3: Further types of generalisations*

Let's assume that the result of executing a program with many individual parts (cellular automata, simulated ants, etc.) is that some of them agglomerate into entities with definite shapes (gliders, a hatchery in an ant colony, etc). The formal reduction argument applies only to the exact state and position of each individual part. As before it may be that the higher level shapes suggested by the individual points are mathematical conjectures, possibly unprovable. Moreover, and especially in the case of an ant colony, the shapes are possibly not exact, or with fuzzy boundaries (see also the controversies about autopoiesis in Section 2.1.1). Yet these would be considered emergent by many definitions, and once again, worth considering in themselves at a higher-level. But as before, they may very well then be formally irreducible. This is now a generalisation in space, not time. When combining both time and space, i.e. when deriving laws for the evolution of the above higher-level entities, then formally irreducible emergence may hold on both aspects. When additionally the simulation is non-deterministic (for example in some cases when using threads or network links, or physical random number generators) the "emergent" entities may very well still be observed. If they do, we'd have an even harder time trying to reduce them to the program formal rules.

*Example 4: Methodological consequences*

We are now given an artificial life simulation, in which we note that on average agents have a preference for doing one kind of action rather than another. Is this statement formally reducible? If we consider the exact runs that were observed and the exact tendency of the agents that was noted, and assuming the simulation is deterministic, yes. But if we want to generalise to other runs, we don't know. Perhaps there are some regions in parameter space where the simulation does not produce this tendency, for example. Reducibility doesn't matter in this case, it's much more fruitful to consider the higher-level in itself (the agents tendency) and apply it a practical approach. By using that "tendency law" and applying it the scientific method as suggested by Jochen Fromm in [FROM06], one may perhaps refine that "tendency law" and find the regions of parameter space where it does not work (if any), or just be satisfied with the law holding for all usual runs. By analogy the Newtonian laws hold for most everyday life situations, though relativistic effects are necessary to explain some observations (and actually may be useful in practice too, for computing GPS positions accurately for example), and neither of them has yet been formally related to the lower-level of particle physics.

### 2.4.3 In practice

The task for understanding a phenomenon, simulated or real, amounts to finding a reasonably precise and concise approximation for that phenomenon and its behaviour. Whether that phenomenon is formally reducible or not cannot be decided generally (see Section 2.3) and does not matter for practical purposes anyway (see above). The difference between formal irreducibility vs formal reducibility is that in the former case the simple shortcut description is necessarily approximative, rather that very often approximative for the formally reducible case (due to the predominance of incompressible statements, see Gregory Chaitin [CHAI74]).

More generally speaking, by considering the higher level entities in themselves (functionalist approach) and trying to formalise their relations directly at the higher level (reductionist approach), one does not need to care whether these relations and entities are "emergent", reducible, or in any way logically connected with the lower-level system, in order to produce satisfying results at the higher level.

Once again, this is reminiscent of what happens in physics: higher level prediction laws (like Newtonian physics) are convenient but imperfect shortcuts for the formal system of equations describing interactions at the nanoscale. The normal procedure is then to try to refine the observations so as to validate or invalidate these laws, potentially leading to the creation of new measuring devices, and so on, until we either improve the higher-level theory or find a better one for explaining the observations. This what Thomas S. Kuhn calls *normal science* [KUHN62]. At this point, downward entailment (see Section 2.2.4 and Russ Abbott's [ABBO06] presentation) may be a way to reconnect the higher-level formal laws with the lower-level system.

The main implication for formal systems and for simulations in particular is that even on a computer, it is possible to observe logically irreducible functionally defined phenomena, thus formally irreducible emergence as previously defined. This is a counter-argument to the formal (logical, causal) reducibility objection to computer simulations: Depending on its setup, a simulation may still be adapted for the study of complex systems and emergence, even the formally irreducible one.

The next section extends on this argument to consider what can actually been done in practice for complex systems.

# 3 Practical investigations

The task of finding a useful approximation usually involves model building and testing. What we can do is collect evidence for micro to macro relationships, and then from these observations try to build a "theory" specific to the system under investigation. The phenomena of interest may not need to be fully and perfectly described by these models. So long as these models and their associated formal systems reasonably explain and allow predictions with quantifiable errors, then we have hope for some degree of control on these emergent phenomena.

Jochen Fromm proposed in [FROM06] that the problem of model building for emergence is equivalent to and should be addressed by the general scientific methodology of theory building, for each phenomena. In other words, each system should be considered as its own little world, with its peculiar rules, and we should try to build theories about higher-level effects in that little world.

Since this is a daunting task, what is proposed here is to concentrate on providing tools and methods that may help in these investigations. There are at least two approaches:

- Consider the new level from scratch. Identify sufficiently stable patterns in system state. In the case where a formal system is present, the identification may rely in part on the dynamical system tools and techniques for detecting attractors. But this is not enough, we also need global tools to identify computationally or formally irreducible effects. A qualitative approach may rely simply on observation (subjective tool). A quantitative one may rely on the premises of a formal system for that new level if such system is available, or more realistically on the detection of repeated patterns (machine learning) in the hope of building such a formal system embryo. In that last case, the goal is to identify the major entities that have a significant role and apply automated tools to build candidate hypothesis of their behaviour. For example, one could start with a crude formal system consisting of the spacial and temporal concentrations of lower-level entities, like agents. This could in turn lead to the identification of groups of agents seemingly "moving" together. This is a first crude concept of a higher-level functional definition for a "group" and an operation "move" on that entity. Then, refinements would consist in defining what exactly that group is, perhaps using pattern matching, what are the laws of motion of that group, group interactions, and so on. Ultimately, a motion theory with formal laws would allow prediction in that system.
- Re-use the observables that already exist to investigate what happens globally, at both the lower and higher levels, and try to relate both. For example, in a recurrent network, a high-level measure could be the learning performance. For an evolutionary experiment a global low-level measure could be the gene diversity. This does not require a formalization of either the lower-level (but it might already exist) and the higher-level (we might have measurements available already). Then, control may be achieved either top-down or bottom-up, by using respectively the global or local measurements to define an objective. In each case, the other measurements provide a way to reach that objective.

Of course, both are complementary and their combination is probably necessary for understanding micro-to-macro relationships. Once a formalism, or at least entities, can be defined at a level of investigation, global measurements may be made on them. For example, the number of prey/predator cycles in an artificial life system relies on the identification of what is a population cycle. We could then monitor and hopefully relate global measurements on the population cycle with global measurements on the agents themselves.

As for the downward entailment example (Section 2.2.4), functional irreducibility splits the micro and macro level formal systems. The aforementioned population cycle measurement refers to the change in time (formal aspect) of a quantity (the population) that is not defined at the level of the agents and their interaction rules (lower-level formal system). It's important to always refer to these precise effects, and not put the "population cycle" concept is the "emergence" bag. The goal is to define the investigation tools that may help refine the emergence concept, not to presuppose what it means beforehand.

Both quantitative and qualitative investigation tools are necessary to understand what happens in a system. The qualitative tools provide a basis for the definition, and further refinements, of concepts and entities acting at a given level. The quantitative tools may help in formalizing a system of "laws" acting on these entities, which in turn is the basis for a "theory". The goal is to be able to perform some prediction and control on the higher-level entities, if only so as to be able to further improve/refute the candidate theory in progress for the system considered.

The next section reviews what could be done for the first point: trying to build and formalize entities from scratch, by investigating data and relations at a given level. Sections 3.2 and 3.3 investigate the second point with the two different approaches: what can be done by the measurement of a global observables to derive micro-macro relationships (global control), and what can be done at the lower-level to relate it to a global property (local control).

## 3.1 Data-based investigation techniques, starting from scratch

### 3.1.1 Presentation

This subsection deals with trying to identify entities and their relations with the following assumption: we only have access to some data. The data is collected beforehand. No mention is done about the formal or functional definitions that lead to the observables that were measured (and which provided the data). The approach is pragmatic: from an unknown lower-lever, data was captured. The goal is in this section to formalize a system at the higher-level consisting of entities and their relations, ultimately forming the basis for a theory at that level.

The term **dynamical regime** is used to refer to some sufficiently stable pattern in system state. The terminology of an attractor would imply some idea of finitude as well as a reference to dynamical systems. A dynamical regime may be transient, or even not associated to particular underlying equations. In an open and dissipative system, this would describe well some sustained pattern, that is not stable in itself, but which is sufficiently persistent so it can be identified, like a whirlpool.

How to identify a dynamical regime? What aspects of the data can be used in the definition of the regime identifier? The approach proposed in this section is to identify dynamical regimes from data only. This is not to say theoretical values can't be used; actually a theory is needed to compile data in a significant way. What is meant is that the regime identifiers are data-driven, in the sense they don't require extra information about an internal or lower-level model. If such information is available for a particular system, then an extension to this method would be to include it, using Baysian statistics for example.

All we have at this point are observed variables (time series) that can be measured. These series should be modified on-line as new data is observed, and older data should be discarded in the case of transient non-stationary systems. Some form of statistics is necessary: By definition, a dynamical regime is observed through time, individual observations are just the limit case. Thus data should be compiled over a time range if a dynamical regime is to be identified correctly. Too short a range implies the risk of not having enough information to identify the regime. Too long a range implies the risk of compiling data from different regimes with non-stationary effects.

In addition, a choice should be made on which observable to apply the analysis on. This choice will also influence the possibility for local control: the result of a control is described in terms of the chosen observable.

Statistics should be interpreted here in the broad sense of the term: a technique to extract some characteristic value from the data, supposedly depending on the dynamical regime. Thus, contrary to basic mean and variance statistics, the data sequence order is important. Actually, when referring to computational irreducibility, what we're looking for may be related to an approximate algorithmic compression of the data: condensing in a few values the information contained in the sequence of values taken by the observables. The argument about formal reducibility or not was treated in the previous theoretical discussion: For all practical purpose, we don't know whether the data is perfectly compressible or not, but this doesn't matter. The goal is to seek for approximate and simpler to comprehend laws and entities. The compression needs not, and should not, be perfect in our case.

Moreover, due to measurement errors, natural variability in the initial conditions, or simply the sampling mechanism, it is expected that a single dynamical regime leads to different time series anyway. Each of these series only have in common the fact they are generated by the same dynamical regime. But each of these series also contains additional information related to the measurement process, sampling mechanism, etc. This additional information should be discarded as it is not related to the dynamical regime itself. Consequently, what we're seeking here is a way to extract some signature, some relevant information, inherent to this particular dynamical regime, and that distinguishes it from other regimes. This identifier should be robust to the aforementioned effects introducing spurious extra information.

All these potential problems call for an intensive but careful use of machine learning techniques. All there is to do[18] is to shift the perspective of machine learning from useful function approximation and optimisation for solving practical problems, to useful tools for the practical investigation of complex systems.

### 3.1.2 Applicability of data-driven techniques

The goal of this document is not to propose a complete overview of the literature about modelling and time series analysis techniques. This was for example proposed by Cosma R. Shalizi in [SHAL06] and Hegger *et al.* in [HKS99]. However, recognizing the issues related to dynamic regime identification is a first step if we intend to solve these problems. In particular, attention should be given to:

- Capturing sequence information. This means finding a way to summarize changes in time of the data, a precondition for identifying time patterns. These patterns relations could then perhaps be used for prediction. For example, times series from an EEG may perhaps lead to the early detection of epilepsy crisis.
- Identifying global trends. This is similar to the seasonal component in weather forecasting. The same measure, like a temperature value of 30°C, will not have the same meaning and predictive impact, according to whether it's in the global trend (ex: in summer, or at the equator) or whether it is anomalous (ex: in winter, or at the poles).

---
18 And without assuming this is an easy task...

- Resistance to noise. As aforementioned the goal is not to obtain a perfect compression of the data but to get robust and simple indicators of what's going on. Noise is expected, the dynamic regime should be robust to small perturbations.
- On-line incremental updating. The goal is to identify dynamical regimes, which must be updated as new data arrives and expired data is discarded. Since dynamic regime identification is only one task that may help in defining observables or formalizing higher-level entities, it is only one step in the full system analysis and must be computationally efficient.
- Maintaining statistical significance, especially when replicating the experiment with different measuring process.
- Finding good observables to apply the techniques to. It's often better to process a synthetic and reliable observable than a thousand noisy data streams. Unfortunately, this is once again a case of circular argument: chances are, supposing there exists an ideal regime identifier, that this identifier would itself be the best observable for other techniques.

Combining different techniques may help capturing different information, like a multifractal analysis (See Muzy *et al.* [MBA93] and Abry *et al.* [AJL04]) provides a different information than a principal component projection (ex: incremental version by Hall *et al.* [HMM98]), a Lyapunov exponent estimation (Hegger *et al.* propose a useful package [HKS99], Rosenstein *et al.* a practical algorithm [RCL93]), etc. Then, an identifier could perhaps be built on top of the various perspectives obtained by different techniques. Global indicators, like the Grassberger-Crutchfield-Young statistical complexity (see Section 2.1.3 and Shalizi [SSH04, SHAL06]), could also be used when they are applicable. In a given context, some combinations may work better than others. Thus it would be interesting to try different regime identifiers in a local to global problem, by checking the influence on the global patterns. Alternatively, the combination of different indicators may simply help in extending the coverage of the identifier. Some techniques fail in particular cases: a time-lag embedding (Sauer *et al.* [SYC91]) may give spurious values due to noise or cycles (Rosenstein *et al.* [RCL94]), a data series may not exhibit a power or an exponential law [HKS99], etc. Combining different techniques could allow an identifier estimation even if one technique fails. The problem with this idea is to ensure the significance of the identifier in the partial failure case. For example, the information "cycle length" is hard to significantly combine with the information "Lyapunov exponent value". In all subsequent usage of these identifiers the distinction remains. Hence, this is not a combination of techniques at all, but rather a list of cases. In turn, this introduces computational artefacts on the observation of higher-level effects based such lists: The different cases may not be significant at higher-level, but they introduce spurious observations. For example, just because a Lyapunov exponent estimation failed due to noise or lack of data doesn't mean there is no attractor at all.

An on-going field of research concerns the development of new data-based mathematical techniques for time series analysis. Combining the existing ones in a computationally efficient, incremental, and easy to apply way is another challenging task. The dynamic regime identifier could be a first step to search for and define higher-level entities. These techniques could also be introduced in machine learning algorithms so as to automate the definition of global observables. These observables and entities could then be used in a second step for the formalization of higher level governing laws, supposing these laws exist for a given system.

## 3.2 Global control: measuring micro-to-macro relationships

This section relies on the assumption a significant observable can be defined globally for a given level of investigation. In the case when a formal system is available for a lower-level, the observable may be defined in that formal system, possibly using techniques from the previous section. In the case some entity can be functionally defined at a higher-level, then measurements may be made on this entity. For example, the global observable could be a pressure, a population count, etc. The working hypothesis is that global measurements are made at a given level. Whether and how the corresponding observable is reducible or not to lower-levels is of no concern for this section.

Global control is necessarily related to one form or another of downward causation, the most effective one being downward entailment, as will be detailed below. A target objective is described using the higher-level observable. The goal of global control is to find the extent of the lower-level states that can produce this higher-level observable.

For example, let's consider autonomous entities in a complex simulation, like prey and predator agents. The previous section has dealt with the possibility to create a higher-level formal system from scratch. For example, we could define a global entity like the agent populations. Let's assume we observe prey-predator cycles. We could then try to derive laws for the evolution in time of the populations, that approximate these cycles. Then, given a current observation, we could perhaps predict according to these laws the future global state of the system, or refine the "population theory" that we're building on that system.

This section deals with the other possibility: creating measurements at global and local levels, and trying to relate both. The population observable does not need a complete higher level formal system to be applicable. It can be used as part of the target objective for global control. Other global measurements may be made: on the population diversity, on the agent's lifespan, etc. A target high-level objective is then defined, for example maintaining a diverse population for as long as possible.

At the lower-level, other observables are defined. For example, the amount of energy that flows in the system, some physical limits for the agents, some possibilities for their interactions, and more generally anything that globally influences the lower-level simulation.

Conceptually, each higher-level observable can take a range of values that depends on the lower-level parameters. In the simplest case, it can take only one value, and this defines a landscape. In the usual case, each higher-level observable, and their combination, may take a restricted range of values depending in a complex way on the lower-level parameter values.

What's missing now is a set of tools that would help relate both worlds. For example, statistics on aggregation values frequently taken by higher-level observables, bifurcation analysis, and more generally anything that may help understanding the higher-level target behaviour, and the shape of the landscape it defines on the lower-level variables in the simplest case.

By analogy with the weak downward causation hypothesis (see section 2.2.4), the global control would then take the form of a restriction of the domain of the lower-level observables. The allowed range would be the one that maximizes the probability for the higher-level target to be reached (ideally with probability 1 for full control). Given a higher-level objective, the range of lower-level possibilities is limited in a way that is not defined at that lower-level (like a population cycle). The only way to understand why these limitations and no other is to consider the higher-level concepts: this is downward explanation as Jaegwon Kim puts it.

The more advanced version of this control is using downward entailment instead of weak downward causation. In this setup, some form of higher-level law has been found that can adequately describe what happens independently of the lower-level, like the hypothetical "population theory" aforementioned, a straight line recurrent pattern, or Newtonian physics to re-use examples from section 2.4.2. Applying this law would produce some constraints at higher level. The second assumption necessary to apply the downward entailment concept is supervenience. This allows to translate the newly found higher-level constraints in lower-level terms. But once again, the range of lower-level possibilities is limited in a way that can only be understood using higher-level considerations.

To summarize:
– Global control is possible without higher level formalization, but such a formalization would help. Reproducible global measurements need only be well-defined, the "reason" for these measurements and the existence of the entities they involve need not be specified.
– Predictions for the higher-level target is made either by restricting the lower-level parameters range directly, or by using a formal high-level "law" acting on the target. In each case, maximizing the prediction reliability is the goal of global control.
– The result is a limitation of the lower-level parameters and rules that may only be understood in terms of a higher-level concept.
– Tools and techniques that help relating both levels play the same role here as tools and techniques for time series analysis and machine learning played in the previous section. Therefore, as before, this document is a call for the creation of a battery of ready-to-use investigations tools.

A simple application of this method in a multi-agent setup was done by the author in [BROD05] with more details in [BROD07A]. That study relied on the gross simplification consisting in averaging the values of the higher-level observables over multiple runs, so as to define a landscape over lower-level parameters. This is a coarse way to study the micro-macro relationships, but nonetheless allows to apply the first global control technique described in this section, and it has provided concluding results.

## 3.3 Local control: engineering lower-levels

Engineering local control rules and running the system is the most conventional way to proceed. Most experiments concerning emergence try to update the local rules and monitor what happens. Usually they then argue whether the global phenomena is expected or not given the assumptions that were made, or whether it can be considered "emergent" or not, which we'll refrain to do given the discussions in section 2. This subsection is therefore not a new approach to the study of complex systems. However, it is the occasion to highlight the change in the way to study them that occurred in the past years.

The dynamical systems point of view is that a higher-level phenomena could possibly be related to internal state attractors. This idea was previously expressed in the discussion about weak downward causation (section 2.2.4). In a recent study Robert Legenstein and Wolfgang Maass [LM07A] show that the "attractor = higher level state" concept is no longer enough. Studies like for example the one by Stefan Bornholdt and Torsten Röhl [BR03] place attractors and their length as just another parameter that changes, and attractor shifts due to noise are an essential part of the investigation. In the context of the previous section, this could even provide an observable for the global state of the system.

In the "large noisy networks with ever-changing attractors view", a dynamical regime would not be associated to one attractor (or sustained cycle), but rather to some global property, with attractors just a component of that global property. It's still possible to use the dynamical regime concept, but it has become less grounded in the attractor or sustained state terminology. More attention is given to the higher-level functional aspect at the expense of the lower-level formal aspect.

The complex systems are still the same, but the way to investigate them has changed. Instead of trying to characterize attractors, Lyapunov exponents, and other dynamical system notions, the attention is now on global statistical properties and measurements. Consequently, the attention shifts from the "far from equilibrium" part to the "thermodynamics" part, to reuse the notions presented in section 2.

An example of this shift of attention is Wolfgang Maass *Liquid State Machine* (LSM) random recurrent spiking neurons network as defined by [MNM03], which has been shown to be able to compute without stable state [MNM02]. A similar approach using sigmoid transfer functions was introduced by Herbert Jaeger as *Echo State Networks* (ESN) [JAEG01]. Both contrast with the previous generation of neural networks, for example multi-layer perceptrons (MLP), where the goal was precisely to make the network converge to a stable state. The LSM and ESN comprise hundred or thousands of nodes, and their recurrent feedback loops are an essential part of the system. The MLP were generally limited to a few tens of nodes, and the feed-forward without recurrent loop aspect was important to prove the convergence analytically. In the LSM setup, the fading memory property [LM07A] assumes the role of the dissipative part of the system. The openness comes from the assumption external energy is available to emit spikes, which by definition are short impulses of energy higher than the rest state. The LSM is then continuously in a sustained state. The MLP were equivalent to a dissipative discrete dynamical system.

Another reason for recent advances seems to be simply the increase of available computing power. When we could only study simple dynamical systems, the precise equation properties, attractors, bifurcations and so on were the main objects of study. Now, all these are acquired parts, not the centre of attention. For example, being able to simulate ten thousand times more nodes in a neural network drastically changes the view of what "dynamic" and "large" means. Some effects are only visible at large scales, like the error rate dropping below a few percents only when the network size is above several hundred nodes [MNM03]. The same way some mathematical properties only appear in higher dimensions, it seems that some properties of the system states only appear in high dimensions. Except for specific particular cases, it's then not practically possible to proceed to a formal analysis.

Local control is still possible, but once again new tools and techniques should be invented or at least existing ones should be adapted. As for the time series and the micro-macro analysis tools aforementioned, this section is a call for qualitative and quantitative tools to statistically describe large scale behaviours. This could be a synthetic parameter, an order/chaos boundary indicator (for example as in Natschläger *et al.* [NBL04], see Section 2.1.3), an average attractor length [BR03], a degree of synchronization, etc.

An example of local control is provided by the author in [BROD07B], with more details given in [BROD07A]: Each node dynamical regime in a LSM is monitored, and then a learning algorithm is derived using these local regime synchronization, in a way reminiscent of Hebbian learning (Song *et al.* [SMA00] version for spiking neurons). Statistical complexity is then computed incrementally in order to quantify the global effect of the local control.

# 4 Conclusion

There is currently no consensus about what the notion of "emergence" entails. The most plausible explanation is that no objective notion of "emergence" matching our intuition can be found, as explained in Section 2.4. Conversely, any comprehensive theory of emergence will have counter-intuitive results.

Thinking in terms of causality and irreducibility helps to clarify the main concepts, as well as to avoid controversies between supporters of different definitions, assumptions and hypothesis that lead to different interpretations. It doesn't seem at this point reasonable or even possible to build a theory of complex systems based only on theoretical arguments.

The conclusion for this analysis is a clear call for the creation of practical investigation tools. It's been proposed that in general the problem of emergence in a given system is equivalent to the problem of building a theory for that system. That is, considering the system as its own little world with its own rules, and trying to formalize laws and entities in this world. Then, using these laws and entities, one could hope to achieve some degree of prediction, if only so as to refine the system-specific theory, and some degree of control, for example through downward entailment.

Such an approach certainly seems to be successful already for some systems like the Game of Life, where a vast community of enthusiasts has created a bestiary of existing entities, what's known about interaction rules, and more[19]. In a sense, the discovery of Turing equivalence for the game of life could not have been possible without first thinking in terms of entities like the gliders and their interactions. The question is, could this be generalized to other frameworks? Could we try to formalize and build a higher-level theory for each complex system? If so, could some common investigation tools and techniques help in this daunting task?

---

19 A good starting point for further investigations is: http://entropymine.com/jason/life/. Example cited from [ABBO06].

In this perspective, the following guideline is proposed:
- Monitor micro-level and macro-level behaviours. Define observables that globally describe each level.
- Using the quantifiers, try to find relations, for example but not only with the help of machine learning techniques. As aforementioned this step is equivalent to theory building for a particular system: there is no magic recipe. This is where reliable investigation tools are an invaluable resource.
- If such micro-to-macro level relationships are found, then there is hope to derive "laws", or shortcuts, that may reasonably provide some degree of prediction and control over emergence. These laws need not be perfect, and in fact in the case of weak and formally irreducible emergence they cannot be so. They need only provide a satisfyingly good approximation for a given problem. In that case, as for any sound scientific theory, refutable predictions should be made with that theory so as to direct further refinements: What Kuhn [KUHN62] calls *normal science*.
- Using either the higher-level formal system embryo or the quantifiers directly, apply some control on the system. If control is sought at the lower level, use the higher-level observables to monitor its effect. This is the most common case. But it's also possible to define functionally a higher-level objective function, and restrict, search, manipulate, or optimize the lower-level parameters so as to reach that objective.

In turn, this raises the question as to how to monitor a system. At this point, it seems reasonable to apply a data-driven approach for the measurements, possibly complemented by the lower-level formal system or the higher-level on-going formalization. One goal is obtaining a direct control, another is to further refine the entities and higher-level formal system, if only to be able to improve the measurements quality, which in turn could lead to more control and predictive ability.

Historically, the development of instruments for measuring the world has been a driving force behind theoretical refinements. There is no reason to think complexity science is different, and better practical tools will lead to better understanding. Although the mathematical development of new tools would be welcome, improving the implementation of these tools and the currently existing ones is a necessity. If ninety percent of the computation time is spent on the measurement process, then there is little room left for applying these instruments to anything in practice. What we need is an efficient and ready-to-use toolbox for making measures, investigating what happens, and help building theories. This can be seen as an help for the scientific formalization of laws and theories about a given system.

This document is therefore a call for a pragmatic, practical approach to complex systems, as well as a call to create generic tools of investigation.

# 5 Acknowledgment

This document presents arguments that were also developed in the author PhD thesis [BROD07A]. See this reference for more details.